\documentclass[12pt]{article}
\usepackage{pic04}
\usepackage{hyperref}
\usepackage{url}
\usepackage{graphicx}

\usepackage{graphicx}  
\usepackage{dcolumn}   
\usepackage{bm}        
\usepackage{amssymb}   

\def\CLB{CL_{B}}
\def\CLS{CL_{S}}
\def\CLSB{CL_{S+B}}
\def\Hpm{ {\it H}^{\pm\pm}}
\def\HpmLR{ {\it H}^{\pm\pm}_{\mathrm L(R)}}
\def\HpmL{ {\it H}^{\pm\pm}_{\mathrm L}}
\def\HpmR{ {\it H}^{\pm\pm}_{\mathrm R}}

\def\Hpp{ {\it H}^{++}}
\def\Hmm{ {\it H}^{--}}

\def\ppbar{\mbox{\it p}\overline{\mbox{\it p}}}
\def\ttbar{\mbox{\it t}\overline{\mbox{\it t}}}
\def\bbbar{\mbox{\it b}\overline{\mbox{\it b}}}
\def\qqbar{\mbox{\it q}\overline{\mbox{\it q}}}
\def\pz{\phantom{0}}

\begin{document}

\title{
Search for Doubly-charged Higgs Boson Production in
the Decay~$\bm{\Hpp\Hmm\to \mu^{+}\mu^{+}\mu^{-}\mu^{-}}$  
with the D{\O} Detector at $\bm{\sqrt{s}=1.96}$~TeV
}
\author{
Marian Zdra\v{z}il     \\
for the D{\O} collaboration.\\
{\em Stony Brook University, Stony Brook, NY 11794-3800} \\}
\maketitle

\baselineskip=14.5pt
\begin{abstract}
A search for the pair production of 
doubly-charged Higgs bosons in the process 
$\ppbar\to\Hpp\Hmm\to \mu^+\mu^+\mu^-\mu^-$ is performed
with the D\O\ Run II detector at the Fermilab Tevatron
using inclusive di-muon events.
These data taken at an energy of $\sqrt{s}=1.96$~TeV
correspond to an integrated luminosity of $113~$pb$^{-1}$
and were recorded by D\O\ between August 2002 and June 2003.
In the absence of a signal,
$95\%$~Confidence Level
mass limits of $M(\HpmL)>118.6$~GeV and $M(\HpmR)>98.1$~GeV 
are set on the pair production cross section 
for left-handed and right-handed
doubly-charged Higgs bosons, assuming $100\%$ branching
into muons and hypercharge $|Y|=2$. 
\end{abstract}

\baselineskip=17pt

\section{Theory}
Doubly-charged Higgs bosons appear in left-right symmetric models, 
in Higgs triplet models~\cite{bib-theo} and in 
little Higgs models~\cite{bib-little}. 
The dominant decay modes are expected to be like-sign lepton pairs,
$\Hpm\to\ell^{\pm}\ell^{\pm}$.
Decay modes with mixed lepton flavor are also possible. In most models, 
the $\Hpm$ coupling to $W$ pairs is suppressed due to the requirement 
that the vacuum expectation value of the neutral member of the Higgs 
multiplet vanishes. This is needed to constrain the $\rho$ parameter 
to unity at tree level.

Pairs of doubly-charged Higgs bosons 
are produced through the Drell-Yan process
$\qqbar\to\gamma^{*}/Z \to\Hpp\Hmm$.
Next-to-leading (NLO) order corrections to this cross section 
have recently been calculated~\cite{bib-spira}. 
The pair production cross sections for left-handed states in the mass
range studied in this analysis are about a factor two larger than for 
the right-handed states due to different coupling to the intermediate 
{\it Z} boson. Left-handed and right-handed states are distinguished by 
their decays into left-handed or right-handed leptons. The cross section 
also depends on the hypercharge $Y$ of the $\Hpm$ boson.

LEP experiments have searched for pair production of doubly-charged 
Higgs Bosons in $e^+e^-$ scattering. 
Mass limits of $M(\HpmL) > 100.5$~GeV 
and $M(\HpmR) > 100.1$~GeV were obtained by OPAL~\cite{bib-OPAL}, 
and a limit of $M(\HpmLR) > 99.4$~GeV by L3~\cite{bib-L3} for
decays into muons. All limits in this Letter are given at $95\%$ 
Confidence Level (C.L.) and for a branching of $100\%$.
Similar mass limits were set for decays into electrons~\cite{bib-OPAL,bib-L3}
or $\tau$-leptons~\cite{bib-OPAL,bib-L3,bib-DELPHI}.

The analysis~\cite{dhpp-prl} is based on inclusive di-muon data taken 
between August 2002 and June 2003. The total integrated luminosity for 
the accepted di-muon triggers is determined to be $113\pm 7$~pb$^{-1}$. 
 
\section{Event selection}
The event selection is performed in four steps. The first step (selection S1) 
requires at least two muons, each with transverse momenta $p_{T}>15$~GeV,
where $p_{T}$ is measured with respect to the beam axis. In the second 
selection (S2), isolation criteria based on calorimeter and tracking 
information are applied to reject background mainly from muons originating 
from semi-leptonic {\it b} quark decays. The next selection (S3) requires 
that for events with exactly two muons the difference in azimuthal angle 
$\Delta\phi$ is less than $4\pi/5$. It is applied to reject $Z\to\mu^+\mu^-$ 
events where one of the charges is mismeasured and to further reduce 
semi-leptonic {\it b} decays. This also removes any remaining small 
background from cosmic muons. In the final selection (S4) at least one 
pair of muons in the event is required to be of like-sign charge. These 
pairs are considered candidates for $\Hpm\to\mu^{\pm}\mu^{\pm}$ decays. 

\begin{table}[htbp]
\caption{\label{tab-like}
Expected number of background events from Monte Carlo, and number of data 
events remaining after each selection cut.}

\begin{center}
\begin{tabular}{lccc}
Selection   & 2 muons & Isolation  & $\Delta \phi<4\pi/5$ \\
(like-sign) & $p_T>15$~GeV         &   &  \\
            & S4 \& S1  & S2 &  S3 \\
\hline  
$Z\rightarrow\mu^+\mu^- $       & $\pz0.9\pz \pm 0.3\pz$   & $0.6\pz\pm 0.2\pz$ & $0.3\pz\pm 0.1\pz$ \\
$ \bbbar $                      & $95.1\pz\pm 3.3\pz$      & $4.4\pz\pm 0.8\pz$ & $0.8\pz\pm 0.2\pz$ \\
$ Z
\to \tau^+\tau^- $   &  $\pz0.6\pz\pm0.3\pz$&  $<0.3$&  $<0.3$ \\
$ \ttbar $                      & $\pz0.24 \pm 0.01$ & $0.11\pm 0.01$ & $0.11\pm 0.01$ \\
$ZZ$                             & $\pz0.06 \pm 0.01$ & $0.05\pm 0.01$ & $0.05\pm 0.01$ \\
$WZ$                            & $\pz0.29 \pm 0.01$ & $0.27\pm 0.01$ & $0.23\pm 0.01$ \\
\hline
MC (sum)                        & $97.2\pz\pm 3.3$   & $5.5\pm 0.7$ & $1.5\pm 0.4$ \\
\hline
data                            & $101$ & $5$ & $3$ \\ 
\\ 
\end{tabular}
\end{center}
\end{table}

When the requirement of having at least one pair of like-sign muons is 
applied at the same time as the selection S1, most of the background
from Z decays is removed, and only 101 like-sign events remain 
(Table~\ref{tab-like}). Since no isolation is imposed
at this stage, the most probable background is due to $\bbbar$
production.

{\sc PYTHIA} is used to estimate this background 
by generating inclusive jet events with
a minimum transverse momentum for the hard interaction
of $30$~GeV~\cite{bib-pythia}. 
The inclusive {\it b} quark production cross section 
$\sigma^{\it b}(p_{\rm T}^{\it b}>30\mbox{~GeV})$ was
measured by D\O\ to be $(54\pm 20)$~nb in the rapidity interval 
$|y^{\it b}|<1$ at $\sqrt{s}=1.8$~TeV~\cite{bib-run1}. 
This cross section, extrapolated to the full $y^{\rm b}$
range and to $\sqrt{s}=1.96$~TeV, is used to normalize the
$\bbbar$ MC sample.

\begin{figure}[htbp]
\begin{center}
\includegraphics[width=0.8\columnwidth]{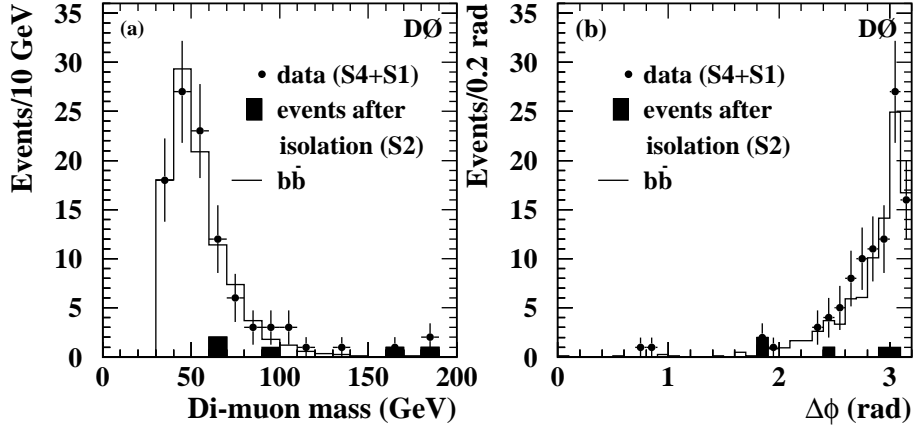}
\end{center}
\caption{\label{fig-like}
(a) Di-muon mass and (b) $\Delta\phi$ distributions for the 101 
like-sign events remaining in data 
after the selections S1 and S4 (points with error bars), compared to 
the {\sc PYTHIA} $\bbbar$ simulation (histogram).
The five events remaining after the isolation selection are shown separately.
}
\end{figure}

In Fig.~\ref{fig-like}, di-muon mass and $\Delta\phi$ 
of the like-sign events are compared to the {\sc PYTHIA} $\bbbar$ simulation. 
Data and Monte Carlo are in good agreement. Out of 101 like-sign events, 
5 remain after applying the isolation requirement, 
while 16 events remain after applying only the $\Delta\phi$ requirement. 
Assuming that all like-sign events originate from $\bbbar$ processes,
the isolation and $\Delta\phi$ selection efficiencies for $\bbbar$
events are $5\%$ and $16\%$, respectively.
Using these the background from $\bbbar$ production
in the final sample is expected to be $0.8\pm0.2$ events.

Another potential background are $Z \to\mu^+\mu^-$ decays where one of the
muon charges is misidentified. We have therefore measured the charge 
misidentification probability in data. $0.45$ background events are expected 
due to charge misidentification. This is in good agreement with $0.3\pm0.1$ 
events expected from Monte Carlo. Three candidates remain in the data after 
the final selection.

\section{Limit Calculation and Systematic Uncertainties}

The limit calculations are performed using the program 
{\sc MCLIMIT}~\cite{bib-tom}. It provides the confidence level for
the background hypothesis, $\CLB$, and the confidence level
for the signal with background hypothesis, $\CLSB$, taking into
account the expected mass distribution for the signal and for
the background, and the mass resolution~\cite{bib-limit}.
The expected signal rate as a function of the Higgs mass 
is given by the NLO cross section~\cite{bib-spira}, 
the signal efficiencies, and the measured luminosity.

The $95\%$ C.L. limit is determined from
the confidence level of the signal
$
\CLS=\CLSB/\CLB
$
by requiring $\CLS=0.05$\footnote{By definition the hypothesis of having a 
signal plus background is excluded at the $95\%$ C.L. if $\CLSB < 0.05$.}.  

The following sources of systematic uncertainty
affecting the normalization of the signal are taken into account:
The systematic uncertainty on the luminosity 
obtained using the D\O\ luminosity system is
estimated to be $6.5\%$. The total uncertainty on the efficiency
amounts to $5\%$. 

The theoretical uncertainty on the NLO $\Hpm$ production cross section from
choice of parton distribution function and variations of the renormalization 
and factorization scales is about $10\%$~\cite{bib-spira}. 

The statistical uncertainty on the MC background rate
is $27\%$ (Table~\ref{tab-like}). Adding the systematic uncertainty of $25\%$ 
on the measured $\bbbar$ cross section~\cite{bib-run1}
yields a total uncertainty on the background rate of $50\%$.

\begin{figure}[htbp]
\begin{center}
\includegraphics[width=0.8\columnwidth]{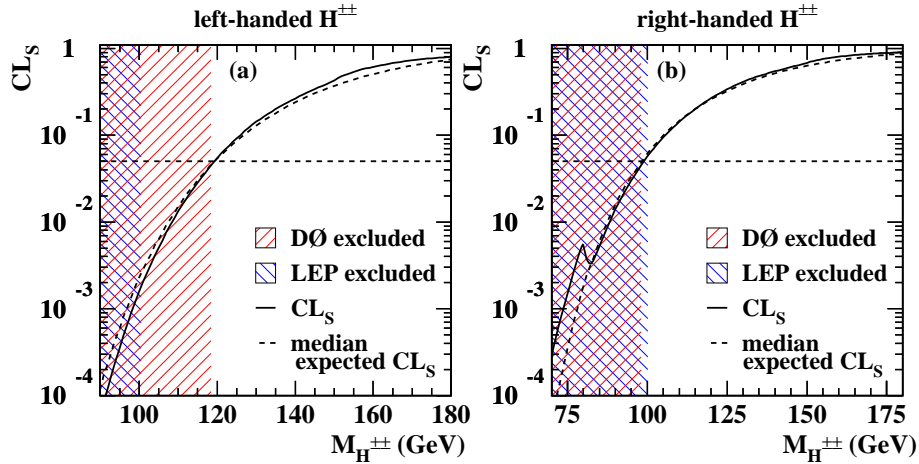}
\end{center}
\caption{\label{fig-cl}
Confidence level of the signal, $\CLS=\CLSB/\CLB$, as a function
of the mass $M(\Hpm)$ of (a) left-handed and (b)
right-handed doubly-charged Higgs bosons.
The mass regions $M(\HpmL)<100.5$~GeV
and $M(\HpmR)<100.1$~GeV are excluded by LEP.
The impact of systematic uncertainties is included in the limits. 
The dashed curve shows median expected $\CLS$ for no signal.
}
\end{figure}

\section{Results}

Taking into account the systematic uncertainties, 
a lower mass limit of $118.6$~GeV
is obtained for a left-handed 
and a mass limit of $98.1$~GeV for a right-handed 
doubly-charged Higgs boson,
assuming $100\%$ branching into muons, hypercharge $Y=|2|$, and
Yukawa couplings $h_{\mu\mu}>10^{-7}$.
This significantly extends the previous mass limit 
of $100.5$~GeV for a left-handed 
doubly-charged Higgs boson~\cite{bib-OPAL}.

\section{Acknowledgments}
The author would like to would like to thank all the people in the
D{\O} Higgs group, namely Stefan S\"{o}ldner-Rembold and Avtanadyl
Kharchilava, for their help.


\begin{thebibliography}{99}

\bibitem{bib-theo}
see, for example: J.F. Gunion, C. Loomis, K.T. Pitts, 
{\em Searching for Doubly Charged Higgs Bosons at Future Colliders},
hep-ph/9610237;
\bibitem{bib-little}
N. Arkani-Hamed et al., JHEP 0208, {\bf 021}, (2002).
\bibitem{bib-spira}
M.~M\"uhlleitner, M.~Spira,
Phys. Rev. D {\bf 68}, 117701 (2003) and private communications.
\bibitem{bib-OPAL}
G. Abbiendi {\it et al.} 
(OPAL Collaboration), 
Phys. Lett. B {\bf 577}, 93 (2003).
G. Abbiendi {\it et al.} 
(OPAL Collaboration), 
Phys. Lett. B {\bf 526}, 221 (2002).
P.D. Acton {\it et al.} 
(OPAL Collaboration), 
Phys. Lett. B {\bf 295}, 347 (1992).
\bibitem{bib-L3}
P. Achard {\it et al.}, 
(L3 Collaboration), 
Phys. Lett. B {\bf 576}, 18 (2003). 
\bibitem{bib-DELPHI}
J. Abdallah {\it et al.},
(DELPHI Collaboration), 
Phys. Lett. B {\bf 552}, 127 (2003). 
\bibitem{dhpp-prl}
Phys. Rev. Lett. {\bf 93}, 141801 (2004)
\bibitem{bib-pythia}
T. Sj\"ostrand, Comp. Phys. Comm. 82, 74 (1994).
\bibitem{bib-run1}
S. Abachi {\it et al.} 
(D\O\ Collaboration), 
Phys. Rev. Lett.  {\bf 74}, 3548 (1995). 
\bibitem{bib-tom}
T. Junk, Nucl. Instruments Methods 
A{\bf 434}, 435 (1999).  
\bibitem{bib-limit}
R. Barate {\it et al.} 
(ALEPH, DELPHI, L3 and OPAL Collaborations and the LEP Working Group
for Higgs Boson Searches), 
Phys. Lett. B {\bf 565}, 61 (2003).
\end{thebibliography}
\end{document}